\documentclass[10pt,twosides,a4paper]{book}
\usepackage{amssymb}
\usepackage[pdftex]{graphicx}
\usepackage{latexsym}
\usepackage[small]{caption2}
\usepackage{amsmath}
\usepackage[english]{babel}
\usepackage{babel}
\usepackage{amsmath,amsthm}
\usepackage{rotating}
\usepackage{multirow}
\usepackage{graphicx}
\usepackage{tocloft}
\usepackage{bbm}
\usepackage{bm}
\usepackage{dsfont}
\usepackage{braket}
\usepackage{amsfonts}
\usepackage{accents}
\usepackage{pdfpages}
\usepackage[thinc]{esdiff}
\usepackage{hyperref}
\usepackage[numbers,sort&compress,square]{natbib}
\usepackage{natbib}
\usepackage{pdfsync}
\usepackage{fancyhdr}
\usepackage{scalerel,stackengine}
\usepackage{fancyhdr}
\usepackage{dsfont}
\renewcommand{\cftchappresnum}{CHAPTER }
\newlength{\mylen}
 \fancypagestyle{plain}{}

\begin{document}

\setcounter{chapter}{6}
\chapter[P. P. Abrantes, D. Szilard and C. Farina: \\
Spontaneous emission and Purcell effect: some aspects]{Spontaneous emission and Purcell effect: some aspects}

\markboth{Spontaneous emission and Purcell effect: some aspects}{P. P. Abrantes, D. Szilard, and C. Farina}

\noindent{\large \textbf{P. P. Abrantes$^{1,2,a}$, D. Szilard$^{1,b}$, and C. Farina$^{1,c}$}}

\vspace{2mm}

\noindent{$^1$Universidade Federal do Rio de Janeiro, 
Rio de Janeiro, RJ, Brazil}\\
\noindent{$^2$Universidade Federal Fluminense, 
Rio de Janeiro, RJ, Brazil}

\noindent{\texttt{$^{a}$ppabrantes91@gmail.com; $^b$daniela@if.ufrj.br; $^c$farina@if.ufrj.br}}

\section{Quantum vacuum effects in the context of QED}

The development of classical electromagnetism resulted from centuries of work by numerous philosophers and scientists, if we take the earliest observations of electrostatic effects in ancient Greece as a starting point. This progress was driven not only by experimental discoveries but also by theoretical advancements culminating in the work of Scottish physicist James Clerk Maxwell. His groundbreaking contributions, encapsulated in his renowned book \lq\lq{\it A Treatise on Electricity and Magnetism\rq\rq}, published in 1873 \cite{Maxwell-1873}, unified Electricity and Magnetism into a single consistent theory. Maxwell's unification could be more deeply appreciated only after the advent of Einstein's special relativity \cite{Einstein-1905}, which establishes how electromagnetic fields transform under changes in inertial reference frames. Beyond this unification, Maxwell's theory also revealed that light was nothing but an electromagnetic wave, integrating optics into the broader framework of electromagnetism. For details on the early history of classical electromagnetism, we recommend the authoritative book written by Whittaker \cite{Whittaker-1951-1953}. The true beauty and elegance of Maxwell's theory is particularly remarkable because, unlike classical mechanics, it did not need to change to accommodate the principles of Einstein's relativity -- it was already consistent with them. One remarkable property of Maxwell's theory is linearity in the absence of sources, which allows the superposition principle to hold. Any linear superposition of solutions to Maxwell's equations is also a valid solution. In our daily lives, this property benefits us enormously -- for instance, the huge amount of electromagnetic waves that simultaneously arrive at our mobile phones do not interact with each other.

Although classical electromagnetism is a general theory with a great variety of applications in macroscopic systems, it has its domain of validity like any other theory. In many scenarios where the relevant length scales are on the atomic or subatomic level, or even in certain macroscopic systems, quantum descriptions are necessary. To date, the most successful theory describing radiation-matter interactions is quantum electrodynamics (QED). Widely regarded as one of the most precise theories ever developed, QED is singled out for its unprecedented agreement between theoretical predictions and experimental observations. A landmark in the construction of QED, often considered its starting point, was Dirac's 1927 paper, in which he established a quantum theory for the emission and absorption of radiation \cite{Dirac-1927}. However, after Dirac's work, all attempts to calculate radiative corrections to electromagnetic processes (such as cross sections) using perturbation theory led to divergent results. These infinite results posed a significant challenge for two decades. Only with Hans Bethe's work \cite{Bethe-1947} on the Lamb shift \cite{Lamb-Retherford-1947} that some light was shed on how to handle these infinities, as evidenced in Dirac's words ({\it apud}~\cite{Milonni-Book-1994} p. 83):

\begin{quote}
{\it
No progress was made for 20 years. Then a development came, initiated by Lamb's discovery and explanation of the Lamb shift, which fundamentally changed the character of theoretical physics. It involved setting up rules for discarding... infinites...}
\end{quote}
In fact, by using a non-relativistic version of QED (atom treated non-relativistically interacting with the quantized electromagnetic field), Bethe successfully calculated the Lamb shift, achieving great agreement with experimental results. Moreover, his approach could be interpreted as a renormalization of the electron mass. Bethe's paper represents another milestone in developing QED, providing a systematic procedure for eliminating spurious and divergent terms in the theory. Another important QED achievement that followed shortly after Bethe’s paper was Schwinger's calculation of the anomalous magnetic moment of the electron (electron $g-2$) \cite{Schwinger-1948}. Schwinger's result matched the electron $g-2$ measurements\footnote{It is worth mentioning that experimental data and theoretical predictions on the electron $g-2$ agree to about one part in a trillion \cite{Beringer-2012}.} made by Kusch and Foley during 1947 and 1948 \cite{Kusch-Foley-1947,Foley-Kusch-1948,Kusch-Foley-1948} (for the early history of the electron $g-2$ see Refs.~\cite{Milena-Dissertacao, Schweber-1994}). 

After that, QED experienced rapid advances and became a theory built on solid foundations. Due to their significant and independent contributions, Tomonaga, Feynman, and Schwinger were awarded the Nobel Prize in 1965. Mention should be made of Dyson, who carried out the task of demonstrating the equivalence of the three formalisms, as clearly outlined in his words \cite{Dyson-1949}:
\begin{quote}
{\it The contribution of the present paper is thus intended to be twofold: first, to simplify the Schwinger theory for the benefit of those using it for calculations, and second, to demonstrate the equivalence of the various theories within their common domain of applicability.}
\end{quote}
For a more detailed history of QED, we strongly recommend the book \lq\lq{\it QED and the Men who Made it\rq\rq} by Schweber~\cite{Schweber-1994}.

If we take the proper limits, QED reduces to Maxwell's classical theory, as expected. In this sense, it is natural to consider the first quantum corrections to Maxwell's theory. Indeed, many interesting and even unexpected phenomena emerge in this semiclassical regime. The first corrections introduced by QED lead to effective Lagrangians that incorporate non-linearities into the theory. The first non-linear terms (quartic terms) that correct Maxwell's Lagrangian appeared in Euler-Kockel's 1935 paper~\cite{Euler-Kockel-1935, Euler-1935}, followed shortly by a more general effective Lagrangian obtained by Heisenberg and Euler \cite{Heisenberg-Euler-1936}. For insightful discussions on effective Lagrangians in QED, see Ref.~\cite{Dittrich-Reuter-1985}. As a consequence, in some situations, the QED vacuum behaves as a non-linear material, capable of being electrically or magnetically polarized by external fields. An example of this phenomenon is the Delbrück scattering \cite{Meitner-Kosters-1933}, in which a photon is elastically scattered by the Coulomb field of a heavy atomic nucleus. This is a remarkable effect because the photon itself has no electric charge to interact with the Coulomb field of the nucleus. This scattering process has already been observed experimentally \cite{JarlskogEtAl-1973, Muckenheim-Schumacher-1980, RulhusenEtAl-1981}. Another surprising effect arising from vacuum polarization is photon splitting, which, like Delbrück scattering, has also been measured \cite{JarlskogEtAl-1973}. A third phenomenon directly resulting from vacuum polarization is light-by-light scattering, first studied by Hans Euler~\cite{Euler-Kockel-1935, Euler-1935} during his doctoral thesis under Heisenberg's supervision\footnote{Hans Euler's thesis originated from a question posed by Peter Debye to Werner Heisenberg in the mid-1930s, when both were professors at the Leipzig University. Debye was wondering whether, within the framework of the newly created quantum theory of electrons and photons by Dirac, light-by-light scattering could occur. Recognizing the relevance of such a question, Heisenberg promptly assigned the problem to his exceptionally talented student, Hans Euler. As W. Dittrich aptly stated \cite{Dittrich-2014}: \lq\lq{\it Hans Euler's Ph.D. thesis is a masterpiece in handling the calculation of low-energy photon-photon scattering.\rq\rq}}.
This process is extremely challenging to observe since the dominant contribution to its cross section is of the order of $\alpha^4$, where $\alpha$ is the fine structure constant. However, after nearly 80 years, the first observations of light-by-light scattering were made by the ATLAS Collaboration \cite{ATLAS-Collaboration-2017}. A comprehensive and pedagogical discussion of this quite unexpected process can be found in Ref.~\cite{Furtado-2019} (in Portuguese).

Another consequence of vacuum polarization is the correction to the Coulomb potential generated by a point charge at rest, commonly referred to as the Uehling potential \cite{Uehling-1935}. In this context, vacuum behaves like a material medium whose polarization gives rise to a screening effect of the point charge. Explicit calculations of such corrections are often presented in standard Quantum Field Theory textbooks for specific distance regimes \cite{Greiner-Reinhart-2010, Peskin-Schroeder-2018}, but exact calculations are also available in the literature \cite{Frolov-Wardlaw-2012, Medeiros-Barone-Barone-2018}.

All the phenomena discussed above are consequences of vacuum polarization. But what happens when vacuum (space without particles and radiation) is subjected to an extremely strong external electromagnetic field? For instance, we know that we cannot charge a conducting sphere indefinitely because, when the electric field at its surface reaches the dielectric rigidity of air (approximately $3\times 10^6$ V/m), small sparks appear, preventing further charge accumulation. A similar phenomenon occurs during atmospheric storms, where lightning flashes provide a dramatic example. The situation is no different in vacuum, except that its \lq\lq{dielectric rigidity\rq\rq} is much greater than that of air under normal conditions. It can be shown that the critical electric field required to destabilize the vacuum is on the order of $10^{18}$ V/m. When the electric field exceeds this threshold, the vacuum becomes unstable, and it becomes energetically favorable for it to decay into particle-antiparticle pairs (mostly electron-positron pairs since the electron is the lightest charged particle in nature). The first discussions of pair creation in an external electric field date back to the 1930s, with the papers by Sauter \cite{Sauter-1931}, who considered the Dirac equation for an electron under a uniform electric field, and Weisskopf \cite{Weisskopf-1936}, who used the concept of the Dirac sea to the QED vacuum. However, a more elegant calculation using a gauge-invariant formalism was carried out by Schwinger in 1951 \cite{Schwinger-1951} (see also Ref.~\cite{Greiner-Muller-Rafelski}). Although electron-positron pairs have not yet been created experimentally using external fields, there is optimism that this may be achieved soon \cite{PairCreation}. For a review of this subject, we recommend Dunne's paper \cite{Dunne-Review}, and for a pedagogical presentation that includes Schwinger's method along with wavefunction techniques, see Holstein's paper \cite{Holstein-1999}. 

However, the QED vacuum is not only susceptible to external fields but also influenced by the presence of bodies such as spheres, plates, or even atoms and molecules. In fact, the Casimir effect, in its standard form, specifically refers to the shift in vacuum energy caused by the presence of two perfectly conducting plates parallel to each other in vacuum\footnote{Vacuum energy shift means a shift in the zero-point energy of the quantized electromagnetic field.} \cite{Casimir-1948}.

Since this energy shift depends on the distance between the plates, it gives rise to a force between them, which is attractive in this case. Nevertheless, the history of the Casimir effect traces back to experiments in colloidal chemistry. After the advent of Quantum Mechanics, Eisenschitz and London \cite{Eisenschitz-London-1930, London-1930} successfully explained the non-retarded dispersive interaction between neutral atoms without permanent multiples but polarizable, known as the van der Waals dispersive interaction. In the 1940s, experiments at the Phillips laboratories focused on colloidal suspensions, and the prevailing theory to explain the equilibrium of such suspensions was based on two competitive interactions: a repulsive electrostatic interaction between colloidal particles due to ion adsorption at their surfaces and the attractive London-van der Waals interaction between them~\cite{Verwey-Overbeek-1948}. However, the theoretical predictions did not match the experimental data. To reconcile the observations, it appeared that the dispersive London interaction should decay faster than the expected $1/r^6$ dependence (with $r$ being the distance between two interacting atoms). Overbeek conjectured that this attenuation of the dispersive interaction could be due to retardation effects, which had not been considered at the time. As Casimir was also working at Phillips, he was asked to investigate this issue. Casimir invited Polder to join him in this non-trivial task, and together they became the first to calculate the retardation effects on the London-van der Waals interaction. Their fourth-order perturbative calculation in QED showed that, at large distances, the dispersion interaction between two atoms decays as $1/r^7$ \cite{Casimir-Polder-1946, Casimir-Polder-1948}. Notably, in these papers, they also analyzed retardation effects on the dispersive interaction between an atom and a perfectly conducting plate. They demonstrated that the difference between the Lennard-Jones non-retarded regime and the retarded one is also characterized by a change in power ($1/z^3 \rightarrow 1/z^4$, where $z$ is the distance between the atom and the conducting plate)\footnote{It seems that Wheeler was the first to suggest that retardation effects in the electromagnetic interaction could weaken the dispersive force between two atoms \cite{Wheeler-1941}.}.
In 1947, Niels Bohr suggested to Casimir that his recent findings with Polder might be related to the zero-point energy of the electromagnetic field. Inspired by this comment, Casimir rederived his results with Polder using a new method, which essentially stated that the dispersion interaction between two bodies is given by the (regularized) variation of the zero-point energy of the electromagnetic field caused by the presence of the interacting bodies. He presented his pioneering findings in May 1948 at a conference on chemical bonds held in Paris, but these results were not published until 1949 \cite{Casimir-1949}. One month later, he presented his famous result on the attraction between two parallel conducting plates in vacuum, published in 1948 \cite{Casimir-1948}. The first experiment to measure this effect was realized by Sparnaay in 1958 \cite{Sparnaay-1958}. Unfortunately, due to poor precision (partly caused by the difficulty in maintaining the plates parallel), Sparnaay could only conclude that \lq\lq{\it The observed attractions do not contradict Casimir's theoretical prediction\rq\rq}. The modern era of precise experiments on the Casimir effect with metals was inaugurated by Lamoreaux \cite{Lamoreaux-1997}, who measured the interaction between a spherical lens and a plate, and by Mohideen and Roy \cite{Mohideen-Roy-1998}, who were the first to use an atomic force microscope to measure the interaction between a sphere and a plate. Since then, many precise experiments using different techniques have been conducted (see Ref.~\cite{Decca-2021} and references therein), enabling us to confidently state that the Casimir effect, one of the most spectacular macroscopic manifestations of the QED vacuum, has been experimentally confirmed with high precision. The Casimir effect \cite{Lifshitz1961, DLP1961} has since become an interdisciplinary topic with applications across various areas of physics, including condensed matter physics, quantum optics, quantum field theory, and cosmology, to name just a few. For pedagogical introductions to the Casimir effect, see Refs.~\cite{Reuter-Dittrich-1985, Elizalde-Romeo-1991, Milonni-Shih-1992, Farina-2006}. For more detailed discussions, we recommend Refs.~\cite{Milonni-Book-1994, Milton-Book-2001, MostepanenkoEtAl-Book, Buhmann-Book}.

Another remarkable phenomenon directly related to vacuum fluctuations is the so-called dynamical Casimir effect\footnote{We underline that Casimir himself never worked on the DCE. The term "dynamical Casimir effect" was coined by Yablonovich in a paper on the Unruh effect~\cite{Yablonovich-1989} and was later popularized by Schwinger~\cite{Schwinger-1983}, who attempted to explain the phenomenon of sonoluminescence using DCE.} 
(DCE). It essentially consists of creating particles from moving (neutral) bodies\footnote{It is worth mentioning that, although DCE is quite a general phenomenon, not every type of motion leads to particle creation.}.
In the context of QED, this means that, for instance, an oscillating conducting plate can generate real photons\footnote{A qualitative way to understand this phenomenon is by considering the interaction between a field and the walls of a cavity with a variable length, which can be thought of as the field interacting with a time-dependent potential. This situation can be compared to a harmonic oscillator (HO) initially in its ground state. If the frequency of the HO starts to vary with time and eventually returns to its original value, it can be shown that the final state of the HO is no longer the ground state but is instead excited into a squeezed state. The parallel here is as follows: the time-dependent frequency of the HO serves as an analogy for the moving walls, the ground state of the HO is analogous to the vacuum state of the field, and the excitation of the HO corresponds to the excitation of the field, meaning real particle creation. For a more pedagogical treatment of the HO with a time-dependent frequency, see Ref.~\cite{Daniel-2020, Daniel-tese}.}.
By simple arguments based on energy conservation, we can conclude that a dissipative force will act on moving bodies, which explains why this phenomenon was originally referred to as \lq\lq{radiation reaction forces on moving boundaries\rq\rq}. The role of the dissipative force is to convert the mechanical energy of the moving body into field energy (real photons). A useful interpretation of this process is that some of the virtual photons, when scattered by the moving boundary, experience frequency shifts large enough to convert them into real photons. The necessary energy for this process is extracted from the mechanical energy of the boundary. This effect was predicted in 1970 by Moore~\cite{Moore-1970}, who considered a one-dimensional cavity with a variable length. Exact solutions for a scalar field with a moving boundary undergoing arbitrary motion (including relativistic cases) were found by Fulling and Davies~\cite{Fulling-Davies-1976}, but in $3+1$ dimensions, only perturbative approaches could be developed. For example, Ford and Vilenkin applied perturbation theory to a scalar field with a Dirichlet boundary condition (BC) at a non-relativistically moving plate~\cite{Ford-Vilenkin-1982}. Further generalizations of Ford and Vilenkin's perturbative approach to the electromagnetic field were made by Maia Neto and his collaborators~\cite{PAMN-1994, PAMN-Machado-1996, Mundurain-PAMN-1998}. The application of the Ford-Vilenkin approach to scalar fields under Robin BC was first performed in Refs.~\cite{MintzEtAl-2006A, MintzEtAl-2006B}. 

However, it can be shown that for non-relativistic moving boundaries, the frequencies of the created photons are smaller than or equal to the dominant frequency of the mechanical movement. Since it is not possible to move material bodies with frequencies greater than a few GHz, the observation of the DCE remains extremely difficult. In fact, it took nearly four decades before this effect was observed for the first time by Wilson {\it et al}~\cite{WilsonEtAl-2011}. This experiment was conducted using an analogue model in superconducting QED, in which the movement of the edge of a superconducting semi-infinite coplanar waveguide terminated by a SQUID was simulated by applying an oscillating magnetic field to the SQUID\footnote{Due to the values of the involved parameters, the time-dependent magnetic flux through the SQUID led to a Robin BC for a scalar field with a time-dependent Robin parameter. Since this parameter has units of length, it is natural to simulate real motion by considering Robin BC with time-dependent parameters, as explored in Ref.~\cite{Silva-Farina-2011}.}~\cite{Johansson-2009}.
DCE continues to be an active research area with many unsolved challenges, such as its microscopic interpretation and the radiation of moving atoms \cite{Reinaldo-2017, MazzitelliEtAl-2019, FoscoEtAl-2021}, the so called asymmetric DCE \cite{JefersonEtAl-2016, JefersonEtAl-2020, AndresonEtAl-2022}, and how to shape dynamical Casimir photons 
\cite{Dalvit-Kort-Kamp-2021}. For more details on DCE, we suggest Refs.~\cite{PAMNEtAl-2011, Dodonov-2020} and references therein.

We finish by mentioning one of the most important phenomena that QED can successfully explain: spontaneous emission (SE). This phenomenon occurs when an excited atom, molecule, or any quantum emitter, even if isolated from all bodies in the universe, inevitably decays to its ground state, emitting one or more photons. Since this is the central topic of this work, we reserve a detailed discussion for the following sections.
%
%
%
%
\section{Spontaneous emission: a historical survey}

Since the main focus of this section is on the history of SE, it is inevitable to mention Einstein's famous 1917 paper~\cite{Einstein-1917}. In this work, he introduced the concepts of spontaneous and stimulated emissions and attributed linear momentum to the quantum of light, the photon\footnote{The name \lq\lq{photon\rq\rq} was not coined by Einstein, but by the chemist Gilbert N. Lewis in 1926. In a letter to Nature  \cite{Gilbert-1926}, Lewis wrote: \lq\lq{\it I therefore take the liberty of proposing for this hypothetical new atom, which is not light but plays an essential part in every process of radiation, the name photon\rq\rq}.}.
The insights contained in Einstein's paper are thoroughly  discussed in Kleppner's work~\cite{Kleppner-2005}, which opens with a remark worth highlighting:
\begin{quote}
{\it The concepts of spontaneous and stimulated emission are well known from Einstein’s 1917 paper on radiation, but his theory of radiation comprises many other concepts - the paper is a treasure trove of physics.}
\end{quote}
One of the key objectives of Einstein's paper was to recover Planck's distribution for a blackbody in thermal equilibrium. Since Planck's seminal work~\cite{Planck-1900}, in which he quantized the energy of harmonic oscillators to derive the spectral distribution of a blackbody at thermal equilibrium, Einstein had become deeply interested in reobtaining this result. In 1916, he wrote a letter to his friend M. Besso in which he stated ({\it apud}~\cite{Milonni-Book-1994} p. 20):
\begin{quote}
{\it A splendid light has dawned on me about the absorption and emission of radiation.}
\end{quote}
Probably, Einstein referred to the ideas of spontaneous and stimulated emissions in his paper. He considered molecules with two energy levels in thermal equilibrium, thereby imposing that the populations in these energy levels should remain constant over time. Using Boltzmann's ideas and making reasonable assumptions about the processes involved -- spontaneous emission, stimulated emission, and absorption -- Einstein ingeniously reobtained Planck's distribution. Although he did not directly compute the SE rate, his calculations allow one to express the ratio between spontaneous and stimulated emissions. Using Einstein's notation, he arrived at
\begin{equation}
    \frac{A_{21}}{B_{21}\rho(\omega_0)} = e^{\hbar\omega_0/K_B T}  -  1\, , 
\end{equation}
where $K_B$ is Boltzmann's constant, $T$ is the absolute temperature, $\omega_0$ is the transition frequency of the molecules, and $\rho(\omega_0)$ is the energy density per unit frequency of the thermal radiation at the transition frequency. Contrary to what one might naively expect, if we consider visible light at room temperature, the above ratio is of the order $10^{39}$, which means that most of the light we see around us comes from spontaneous emission.

The first person to compute the SE rate of an excited atom in vacuum was Dirac in 1927, obtaining (we are writing Einstein's $A_{21}$ coefficient as $\Gamma^{(0)}$) \cite{Dirac-1927}
\begin{equation}
    \Gamma^{(0)} = \frac{4}{3} \frac{|{\bm d}|^2 \omega_{0}^3}{\hbar}\, , \label{TaxaEEEL}
\end{equation}
where ${\bm d}$ is the corresponding transition dipole moment. For many years, this result was considered an intrinsic property of each atom, meaning it should not depend on the presence of bodies near the quantum emitter, such as a conducting wall. After all, how could the atom become aware of the wall before emitting the photon \lq\lq to see{\rq\rq} the wall.

Around two decades after Dirac's result, Purcell demonstrated that the environment can influence the SE rate of an atomic system. As Purcell put in his 1946 work \cite{Purcell-1946}:
\begin{quote}
\it ... However, for a system coupled to a resonant electrical circuit, the factor $8\pi\nu^2/c^3$ no longer gives correctly the number of radiation oscillators per unit volume, in unit frequency range, ... The spontaneous emission probability is thereby increased, and the relaxation time reduced, by a factor $ {f = 3Q\lambda^3/4\pi^2 V}$, where $V$ is the volume of the resonator.
\end{quote}
The excited atom does not need to emit a photon to see the wall. The presence of the wall (or any other object) imposes BCs on the field modes with which the atom interacts, changing its SE rate. This phenomenon is known as the Purcell effect.

%
%
%
%
\section{Spontaneous emission rate: simple examples}

This section calculates the SE rate for a few simple situations. We begin by presenting a general formula for the SE rate expressed in terms of the electromagnetic field modes that satisfy appropriate BCs. For simplicity, we consider a two-level quantum emitter, but generalizations to a multilevel emitter are straightforward.  

\subsection{Spontaneous emission rate}

The system under consideration consists of an atom (or molecule, or any other quantum emitter) and the quantized electromagnetic field, submitted to the BC imposed by the bodies in the neighborhood of the atomic system. We adopt a perturbative approach, where the basis of states is the tensor product of the atomic states and the electromagnetic field states in the absence of the atom, but subject to the appropriate BC. In the initial state $\vert i\rangle$, the atom is in its excited state, and the field is in its ground state. In the final state $\vert f\rangle$, the atom is in its ground state, and the field is in a state with one photon. 

In the dipole approximation, the Hamiltonian of the system is given by ${\hat{{\cal H}}_A + \hat{{\cal H}}_F + \hat{{\cal H}}_{I}}$, where $\hat{{\cal H}}_A$ represents the atomic Hamiltonian, $\hat{{\cal H}}_F$ is the field Hamiltonian, and the interaction Hamiltonian is given by
\begin{equation}
\hat{{\cal H}}_{I} =  - \hat{{\bm d}}\cdot \hat{{\bm E}}({\bm r}), \,
\end{equation}
with $\hat{\bm d}$ being the atomic dipole operator and ${\hat{\bm E}}({\bm r})$ the electric field operator at the atom's position.

From Fermi's golden rule, the desired rate is given by
\begin{equation}
\Gamma({\bm r}) = \dfrac{2\pi}{\hbar^2}\sum_{\{f\}}|\langle i|\hat{{\cal{{H}}}}_{I} |f\rangle|^2 \delta(\omega-\omega_{0})\, .
\label{RegraDeOuroDeFermi}
\end{equation}
A straightforward calculation yields~\cite{Yuri-Tese, tese-Dani}
\begin{align}
\Gamma ({\bm r}) = \frac{\pi \omega_0}{\epsilon_0 \hbar} \sum_\zeta |{\bm d} \cdot {\bm A}_\zeta ({\bm r})|^2 \delta (\omega_\zeta - \omega_0) \, ,
\label{TaxaEE}
\end{align}
where the Coulomb gauge was assumed,  $\nabla\cdot {\bm A}_{\zeta}({\bm r}) = 0$,
the field modes  ${\bm A}_\zeta ({\bm r})$ are solutions  of the Helmholtz equation, 
$(\nabla^2 + k_{\zeta}^2){\bm A}_{\zeta}({\bm r}) = 0$, and satisfy the appropriate BC. Additionally, they form an orthonormal set, such that $\int d^3{\bm r}\, {\bm A}_{\zeta}^{*}({\bm r}) \cdot {\bm A}_{\zeta'}({\bm r}) = \delta_{\zeta \zeta'}$.

A few comments are in order. First, Eq.~(\ref{TaxaEE}) is exactly the same as that obtained via the Weisskopf-Wigner theory
\cite{Weisskopf-Wigner-1930} for the SE rate of a two-level quantum emitter influenced by bodies of arbitrary shapes in its neighborhood. In 1930, they proposed that the probability for the system to remain in its initial state could, to a good approximation, decay exponentially. In our framework, this assumption corresponds to what is called the Markovian approximation, which states that the system has no memory of prior instants. Secondly, we assume that the transition dipole moment is not zero. In case ${\bm d}$ vanishes due to selection rules, a higher order calculation is required, which would lead to two-photon spontaneous emission (TPSE). For more details on TPSE in various contexts, see Refs.~\cite{GoppertMayer, Yuri-Tese} and references therein. Lastly, note that the previous equation explicitly depends on the electromagnetic field modes, which, in turn, are influenced by the presence of nearby bodies due to the BCs imposed by these bodies. Hence, it becomes evident that the SE rate of a quantum emitter is not an intrinsic property, but rather depends on the surrounding atomic environment. With the above reasoning, the Purcell effect emerges as a quite obvious and expected phenomenon.

%
\subsection{Atom in free space}
In the absence of BCs, the normal modes are plane waves, given by~\cite{Milonni-Book-1994}
\begin{equation}
    {\bm A}_{{\bm k}p}({\bm r}) = \frac{e^{i{\bm k}\cdot {\bm r}}}{\sqrt{V}} {\bm e}_{{\bm k} p} \,,
\end{equation}
\noindent where ${\bm e}_{{\bm k} p}$ is the polarization vector for each mode with wavevector ${\bm k}$ and polarization $p$. By incorporating this result into Eq.~(\ref{TaxaEE}), transitioning to the continuum according to $\sum_\zeta \rightarrow \sum_{{\bm k} p} \rightarrow V/(2 \pi)^3 \sum_{p = 1, 2} \int d^3 {\bm k}$, and calculating the integral in spherical coordinates, we obtain exactly Eq.~(\ref{TaxaEEEL}). As expected, $\Gamma^{(0)}$ does not depend on the atom's position since the space is isotropic and homogeneous. Typical excited-state lifetimes $\tau = 1/\Gamma^{(0)}$ for optical electric dipole transitions range from $10^{-7}$ to $10^{-9}$ s. For example, in the case of the hydrogen atom's $2p \rightarrow 1s$ transition, $\tau \sim 2 \times 10^{-9}$ s.

As we will see in the following sections, the presence of objects breaks translational symmetry, making the SE rate dependent on the atom's position.
\subsection{Atom near an infinite conducting plate}
As a second example, we analyze the influence of an infinite, perfectly conducting plate at $z = 0$ on the SE rate of an atom placed at a distance $z$ from the plate. The field modes in the half-space $z > 0$ that satisfy the Helmholtz equation and the BC ${\bm E_{\parallel}}({\bm r},t){\big|}_{z=0} = {\bm 0}$ (which implies ${\bm A}_{{\bm k}p}^{\parallel}({\bm r}){\big|}_{z=0} = {\bm 0}$) are~\cite{Milonni-Book-1994}
\begin{align}
    {\bm A}_{{\bm k}1}({\bm r}) &= \sqrt{\frac{2}{V}} (\hat{{\bm k}}_{\parallel} \times \hat{{\bm z}}) \sin(k_z z) e^{i{\bm k}_{\parallel}\cdot {\bm r}}\, , \label{ModosPlacas1}\\
    {\bm A}_{{\bm k}2} ({\bm r}) &= \frac{1}{k} \sqrt{\frac{2}{V}} \left[k_{\parallel}\cos(k_z z)\hat{{\bm z}} - ik_z\sin(k_z z)\hat{{\bm k}}_{\parallel}  \right] e^{i{\bm k}_{\parallel}\cdot {\bm r}} \,. \label{ModosPlacas2}
\end{align}
\noindent Here, we have decomposed ${\bm k}$ into its parallel component ${\bm k}_\parallel = k_x \hat{\bm x} + k_y \hat{\bm y}$ and perpendicular component $k_z \hat{\bm z}$ with respect to the plate.

It is convenient to write ${\bm d} = d_\perp \hat{\bm z} + {\bm d_\parallel}$ to investigate the quantum emitter's lifetime when suitably prepared such that its transition dipole moment is either perpendicular or parallel to the surface. Inserting Eqs.~(\ref{ModosPlacas1}) and (\ref{ModosPlacas2}) into Eq.~(\ref{TaxaEE}) leads to~\cite{Milonni-Book-1994}
\begin{align}
	\frac{\Gamma_\perp (z)}{\Gamma^{(0)}} &=  1 - \frac{3 \cos (2 k_{0} z)}{(2 k_{0} z)^2} + \frac{3 \sin (2 k_{0} z)}{(2 k_{0} z)^3} \,,\\
	\frac{\Gamma_\parallel (z)}{\Gamma^{(0)}} &= 1 - \frac{3 \sin (2 k_{0} z)}{2 (2 k_{0} z)} - \frac{3 \cos (2 k_{0} z)}{2 (2 k_{0} z)^2} + \frac{3 \sin (2 k_{0} z)}{2 (2 k_{0} z)^3} \,,
\end{align}
\noindent with $k_{0} = \omega_{0}/c$, and $\Gamma_\perp$ ($\Gamma_\parallel$) represents the contribution of ${\bm d}_\perp$ (${\bm d}_\parallel$) to the decay rate. When all orientations of the transition dipole moment are equally probable, the SE rate can be obtained by summing the previous contributions, taking the orientational average, which yields $\Gamma_\textrm{iso} = \Gamma_\perp/3 + 2\Gamma_\parallel/3$.

\begin{figure}[t]
\centering
\includegraphics[scale=0.45]{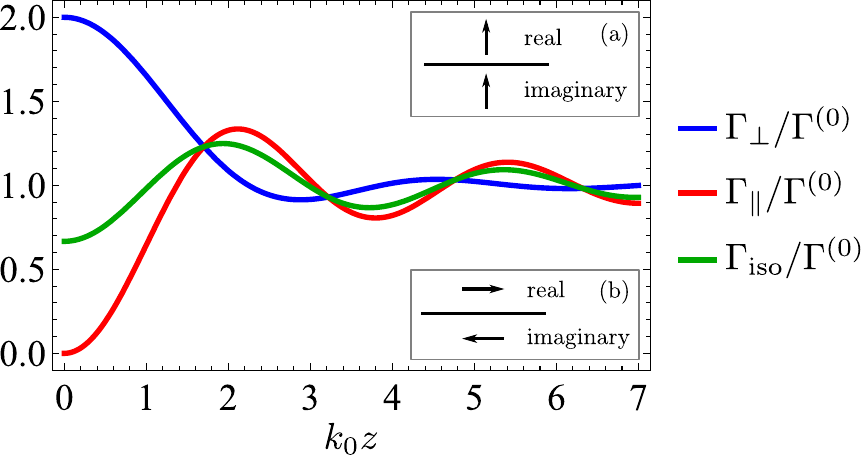}
\caption{SE rates $\Gamma_\perp/\Gamma^{(0)}$, $\Gamma_\parallel/\Gamma^{(0)}$, and $\Gamma_\textrm{iso}/\Gamma^{(0)}$ as functions of $k_{0} z$. Insets (a) and (b) illustrate the classical analogues of a quantum emitter with the transition dipole moment perpendicular or parallel to the perfectly conducting plane.
}
\label{Fig-Atom-Plate}
\end{figure}

Figure \ref{Fig-Atom-Plate} shows $\Gamma_\perp/\Gamma^{(0)}$, $\Gamma_\parallel/\Gamma^{(0)}$, and $\Gamma_\textrm{iso}/\Gamma^{(0)}$ as functions of $k_{0} z$. As $k_{0} z \rightarrow 0$, $\Gamma_\parallel \rightarrow 0$, $\Gamma_\perp \rightarrow 2 \Gamma^{(0)}$, and $\Gamma_\textrm{iso} \rightarrow 2\Gamma^{(0)}/3$. These results can be interpreted using the image method, where the quantum emitter is modeled as a classical radiating dipole (see insets in Fig.~\ref{Fig-Atom-Plate}). In the perpendicular configuration, the real and image dipoles add up, effectively doubling the dipole moment. In contrast, in the parallel configuration, the real and image dipoles oscillate in opposite phases, canceling each other and suppressing the dipole radiation. Additionally, for $k_{0} z \rightarrow \infty$, the SE approaches the free-space result, as expected. Intermediate distance regimes lead to oscillations in the SE rates, caused by interference between the two contributions to the radiation reaction field at the atom's position: one contribution that would already exist in the absence of the plate, and the other due to the reflection of the first one on the plate. Finally, note that $\Gamma_\parallel/\Gamma^{(0)}$ displays stronger oscillations than $\Gamma_\perp/\Gamma^{(0)}$. This pattern appears because transverse electric modes do not couple with electric dipoles perpendicular to the metallic plate. Therefore, ${\bm d}_\parallel$ interacts with both transverse electric and magnetic modes, whereas ${\bm d}_\perp$ couples only with transverse magnetic modes. This effect on atomic lifetime was experimentally verified by Drexhage, Kuhn, and Shäfer~\cite{Drexhage-66, Drexhage1968}.

\subsection{Atom near two parallel and infinite conducting plates}\label{Atom-2Plates}
Consider the atom at position $z$ between two parallel, infinite, and perfectly conducting plates located at $z = 0$ and $z = L$. The field modes in the region $0 \leq z \leq L$ that satisfy the Helmholtz equation and the BCs at each surface (${\bm E_{\parallel}}({\bm r},t){\big|}_{z=0} = {\bm E_{\parallel}}({\bm r},t){\big|}_{z=L} = {\bm 0}$) are, once again, given by Eqs.~(\ref{ModosPlacas1}) and (\ref{ModosPlacas2}), but with $k_z = m\pi/L$ ($m = 0, 1, 2,...$).

By carrying out the calculations for the SE rates for the parallel and perpendicular contributions, we obtain~\cite{Milonni-Book-1994}
\begin{align}
    \frac{\Gamma_{\parallel}(z)}{\Gamma^{(0)}} &= \frac{3\lambda_{0}}{4L} \sum_{m=1}^{N} \left(1+\frac{m^2\lambda_0^2}{4L^2}\right)\sin^2\left(\frac{m\pi z}{L}\right)\, , \label{EEParallel2placas} \\
    \frac{\Gamma_{\perp}(z)}{\Gamma^{(0)}} &= \frac{3\lambda_{0}}{4L} \left[1+\sum_{m=1}^{N} 2\left(1-\dfrac{m^2\lambda_0^2}{4L^2}\right)\cos^2\left(\frac{m\pi z}{L}\right)\right] \,, \label{EEPerp2placas}
\end{align}
\noindent where $N$ is the largest integer less than or equal to $2L/\lambda_0$. Note that, when $L < \lambda_{0}/2$, the SE rate for the parallel configuration is completely suppressed. However, this suppression does not apply to $\Gamma_\perp$, as the sum in Eq.~(\ref{EEPerp2placas}) starts at $m = 0$ and is non-zero.

\begin{figure}[t]
\centering
\includegraphics[scale=0.45]{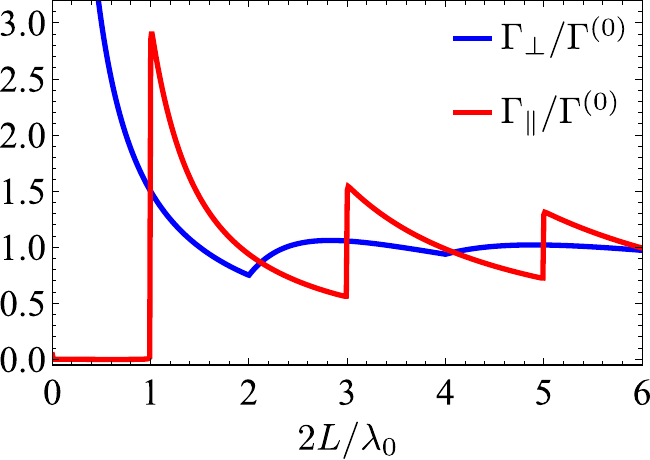}
\caption{SE rates $\Gamma_\perp/\Gamma^{(0)}$ and $\Gamma_\parallel/\Gamma^{(0)}$ as functions of $2L/\lambda_{0}$ with the atom placed at $z = L/2$.}
\label{Fig-Atom-2Plates}
\end{figure}

These two scenarios are illustrated in Fig.~\ref{Fig-Atom-2Plates}. As the plates move closer, previously allowed electromagnetic field modes no longer satisfy the BCs, causing discontinuities in the SE rate for the parallel configuration. This continues until the separation between the plates reaches $L = \lambda_0/2$, when only a single mode remains allowed. For $L < \lambda_0/2$, no mode that couples with ${\bm d}_\parallel$ remains, and the SE rate for the parallel configuration is entirely suppressed. 
This effect was experimentally observed by Hulet, Hilfer, and Kleppner~\cite{Hulet-85}.

%
%
%
%
\section{Purcell effect in some non-trivial situations}
Over the past decades, advances in nano-optics, plasmonics, and metamaterials have led to significant progress in controlling SE. Researchers have employed intricate nanostructures, such as optical cavities, photonic crystals, metallic nanostructures, and plasmonic devices, to manipulate the local density of states (LDOS) and enhance light-matter interactions. From now on, we will explore several non-trivial scenarios in which the presence of surrounding bodies and external agents can greatly alter the SE rate.
\subsection{Atom near a plasmonic cloaking device}
In the last section, we discussed the Purcell effect, illustrating how the surrounding environment and the orientation of the transition dipole moment of the emitter can either enhance, attenuate, or even suppress the SE rate. Another possibility of suppression occurs with an atom inside a photonic crystal\footnote{A photonic crystal is a medium with a periodic dielectric constant, leading to forbidden wave propagation in certain frequency ranges, called photonic bandgaps. This behavior is similar to electron waves propagating through a crystal lattice. For more details on photonic crystals, see Refs.~\cite{JoannopoulosEtAl-1997, JoannopoulosEtAl-2008}.}.
If the atomic transition frequency lies inside the photonic bandgap, no field mode will be available to couple with the atom, effectively suppressing the SE \cite{Yablonovich-1987}. 
 
In the previous examples, total suppression occurred for the SE rate. However, if we aim to suppress only the Purcell effect caused by a surrounding body, such as a dielectric sphere, we face the question: Can we make the sphere \lq\lq{invisible\rq\rq} to the atom, so its SE rate matches that of an atom in free space?

The answer is yes! The key is to ask first whether it is possible to render a dielectric sphere invisible when illuminated by (classical) radiation at a specific frequency. In other words, can we create a mechanism to minimize or eliminate the scattered radiation by the sphere? This mechanism, known as plasmonic cloaking, was proposed by Al\`u and Engheta in 2005 \cite{Alu-2005, Alu2-2005}. It involves a scattering cancellation mechanism based on the unique properties of negative polarizabilities in plasmonic materials and some metamaterials. The idea is that the radiation scattered by the cloak strongly reduces the scattering from the object being cloaked (at a specific frequency). Invisibility is measured by the decrease in the total scattering cross section $Q_s$. 

\begin{figure}[h!]
  \centering
\includegraphics[width=0.85\linewidth]{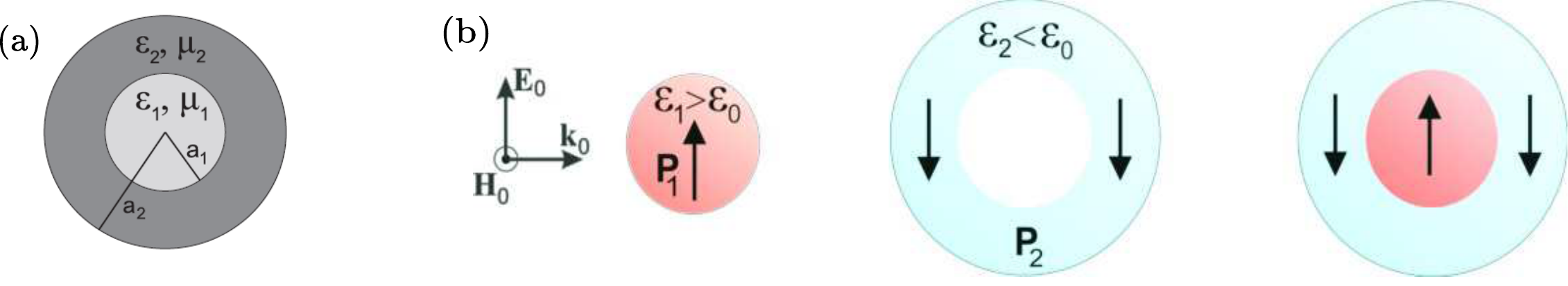}
  \caption{(a) Dielectric sphere of radius $a_1$, permittivity $\varepsilon_1$, and permeability $\mu_1$ covered by a spherical shell of inner radius $a_1$ and outer radius $a_2$, permittivity $\varepsilon_2$, and  permeability $\mu_2$. \linebreak (b) Schematic diagram representing the cancellation of the core sphere and outer shell polarizations' when $\varepsilon_2 < \varepsilon_0$, where $\varepsilon_0$ is the vacuum electric permittivity. 
  Adapted from Refs.~\cite{Kort-Kamp-2013, tese-Wilton}. }
  \label{Cloak-sys}
\end{figure}

Consider a dielectric sphere of radius $a_1$, electric permittivity $\varepsilon_1(\omega)$, and magnetic permeability $\mu_1(\omega)$, covered by a spherical shell with inner radius $a_1$ and outer radius $a_2$, made of a plasmonic material with electric permittivity $\varepsilon_2(\omega)$ and magnetic permeability $\mu_2(\omega)$, as shown in Fig.~\ref{Cloak-sys}(a). For convenience, we consider non-magnetic materials, so $\mu_1(\omega) = \mu_2(\omega) = \mu_0$, where $\mu_0$ is the vacuum magnetic permeability.

For an object with spherical symmetry, the Mie coefficients $c_{n}^{TM}$ and $c_{n}^{TE}$ relate the scattered fields to the incident ones  
\cite{BohrenHuffman-book, Chew-book}. It can be shown that the total cross section is~\cite{BohrenHuffman-book}
\begin{equation}
    Q_s = \frac{2\pi}{|k|^2} \sum_{n=1}^{\infty} (2n+1) \left(\left|c_n^{TE}\right|^2 + \left| c_n^{ TM} \right|^2 \right) \,. \label{Qs}
\end{equation}
For large objects, many terms of Eq.~(\ref{Qs}) will significantly contribute to $Q_s$. However, for small ones ($d\ll\lambda$, where $d$ is the object's size and $\lambda$ is the wavelength of the impinging radiation), the electric dipole moment dominates (corresponding to the Mie coefficient $c_1^{TM}$). In what follows, we assume this last condition is valid and consider only the electric dipole contribution.

Covering the inner sphere with a shell of the same material enlarges the sphere and increases the total scattering cross section. However, an appropriate choice of $\varepsilon_2$ can strongly reduce $Q_s$. For instance, if $\varepsilon_2 < \varepsilon_0$, the polarization on the shell will oppose that on the sphere, as illustrated in Fig.~\ref{Cloak-sys}(b). By carefully selecting the thickness of the plasmonic shell, we can achieve a situation in which the total electric dipole of the system (sphere + plasmonic shell) vanishes, thereby suppressing the total scattering cross section. While the suppression is never complete due to the contributions of the higher multipoles, for $d\ll\lambda_0$, if $c_1^{TM} = 0$, the total cross section will be extremely small. The invisibility condition in the dipole approximation means precisely that $c_1^{TM}=0$. 

\begin{figure}[t]
  \centering
 \includegraphics[width=\linewidth]{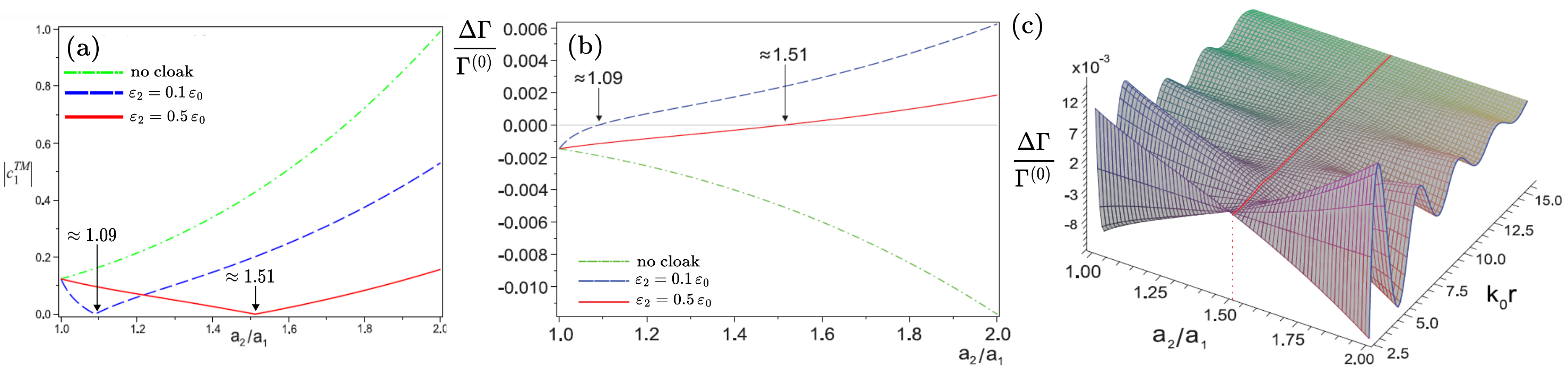}
  \caption{(a) $|c_1^{\rm TM}|$ as a function of $a_2/a_1$. (b) Relative SE rate $\Delta\Gamma({\bm r})/\Gamma^{(0)}$ as a function of $a_2/a_1$ for the atom-sphere distance $k_0 r_0 = 10$. (c) $\Delta\Gamma ({\bm r})/\Gamma^{(0)}$ as a function of both $a_2/a_1$ and $k_0 r$, with $\varepsilon_2 = 0.5 \varepsilon_0$. In all plots, $a_1 = \lambda_0/10$ and $\varepsilon_1 = 10\varepsilon_0$. Adapted from Ref.~\cite{Kort-Kamp-2013}.}
  \label{Cloak_results}
\end{figure}

Figure~\ref{Cloak_results}(a) illustrates this situation. The green line shows the case where the spherical shell has the same permittivity as the inner sphere, meaning no cloak, which causes $Q_s$ to increase with $a_2/a_1$. The red line corresponds to $\varepsilon_2 = 0.5\, \varepsilon_0$, which is smaller than $\varepsilon_0$, thereby existing a value of $a_2/a_1$ such that $c_1^{TM}=0$. The blue line shows the result for $\varepsilon_2 = 0.1\, \varepsilon_0$, where the smaller permittivity leads to a thinner plasmonic shell needed to satisfy the invisibility condition ($c_1^{TM}=0$) compared to the case with $\varepsilon_2 = 0.5 \, \varepsilon_0$.

Let us return to the original question of whether the Purcell effect can be canceled. Kort-Kamp {\it et al} showed that, within the dipole approximation, this is indeed possible \cite{Kort-Kamp-2013} (see also Ref.~\cite{tese-Wilton}). By calculating the scattered field modes for this cloaking device, they used Eq.~(\ref{TaxaEE}), concluding that the suppression of the Purcell effect occurs exactly under the same invisibility condition described above for classical radiation. 

Figure~\ref{Cloak_results}(b) presents the relative variation of the SE rate normalized by its free-space value, $\Delta\Gamma({\bm r})/\Gamma^{(0)} = [\Gamma({\bm r}) - \Gamma^{(0)}]/\Gamma^{(0)}$, as a function of $a_2/a_1$ for a fixed emitter-sphere distance $k_0r_0 = 10$. The inner sphere has permittivity $\varepsilon_1 = 10\, \varepsilon_0$ and radius $a_1 = \lambda_0/10$, consistent with the dipole approximation. This figure compares cases with and without a cloak. Without the cloak (green line), as the sphere's radius increases, the SE rate deviates more from the free-space value, as expected. In contrast, with the plasmonic shell in place, $\Delta\Gamma ({\bm r})$ approaches zero ($\Gamma ({\bm r}) \rightarrow \Gamma^{(0)}$), indicating suppression of the Purcell effect as the cloak’s outer radius $a_2$ increases. More notably, the SE rate remains identical to its free space value for $\varepsilon_2 = 0.1 \, \varepsilon_0$, when $a_2/a_1 = 1.09$, and for $\varepsilon_2 = 0.5\, \varepsilon_0$, when $a_2/a_1 = 1.51$, corresponding to the classical invisibility condition.

Figure~\ref{Cloak_results}(c) presents a 3D plot of $ \Delta \Gamma ({\bm r})/\Gamma^{(0)} $ as a function of both $a_2/a_1$ and $k_0 r$. The inner sphere parameters are the same as in Fig.~\ref{Cloak_results}(b), with $\varepsilon_2 = 0.5\,\varepsilon_0$. The SE rate exhibits oscillatory behavior with respect to $k_0 r$, except at $a_2/a_1 \simeq 1.51$ (highlighted by the red line), where the invisibility condition holds. Note that the SE rate remains equal to its free-space value $\Gamma^{(0)}$ across all values of $k_0 r$. However, when $a_2/a_1$ exceeds the invisibility condition, the shell’s contribution to the scattered field becomes dominant, making the system visible and causing the relative SE rate to increase.

As a final comment, we emphasize that the emission properties of an atom can be used to locally probe the efficiency of plasmonic cloaking devices instead of explicitly calculating the total scattering cross section of the object. 
\subsection{Effects of a percolation phase transition in the SE rate}
Critical phenomena and phase transitions are key topics in physics, influencing structural, thermal, and electrical properties of materials \cite{Stanley-99, Zheng-2022}. Composite media provide a versatile platform for achieving unconventional material behaviors, including phase transitions and pronounced LDOS fluctuations~\cite{Huang-2006}. At scales much larger than their inhomogeneities, the electromagnetic and optical responses of these media can be described using homogenization techniques, such as the Bruggeman effective medium theory (BEMT)~\cite{BEMT}, allowing one to determine an effective dielectric constant based on the properties, shapes, and volume fractions of the constituents.

\begin{figure}[t]
  \centering
 \includegraphics[width=0.42\linewidth]{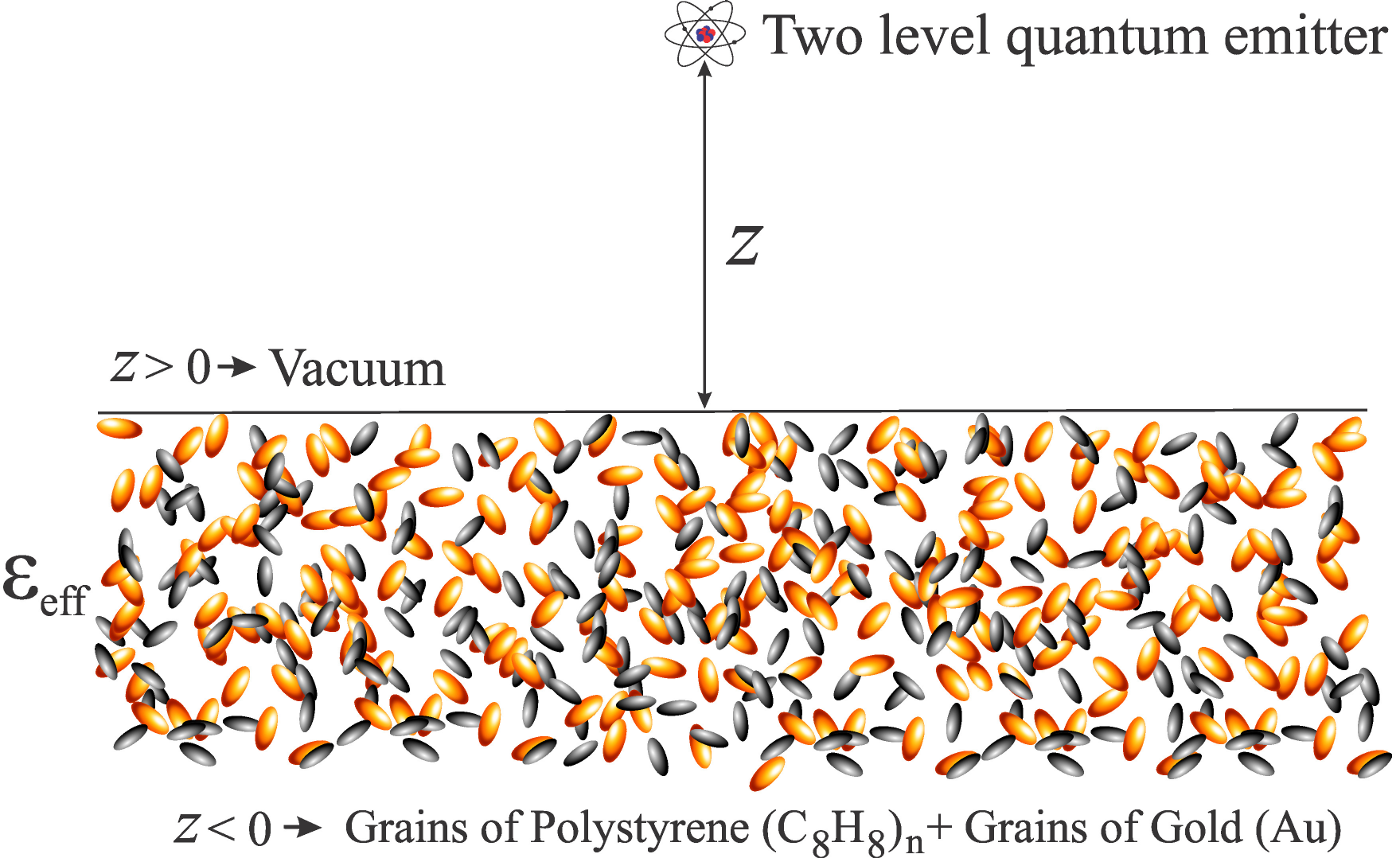}
  \caption{Quantum emitter close to a composite medium, formed by randomly distributed metallic inclusions embedded in a dielectric host matrix. From Ref.~\cite{Szilard-2016}. }
  \label{Percolation}
\end{figure}

In Refs.~\cite{Szilard-2016, tese-Dani}, we investigate the SE rate of an emitter at a distance $z$ of a semi-infinite medium composed of randomly distributed metallic inclusions embedded within a dielectric host matrix, as shown in Fig.~\ref{Percolation}.

So far in this chapter, we have used Eq.~(\ref{TaxaEE}) in terms of electromagnetic modes to calculate SE rates. However, in cases where calculating these electromagnetic modes becomes more involved, it is convenient to adopt the Green's function formalism. In the presence of an arbitrary environment, the SE rate can be obtained from \cite{Novotny-Book}
\begin{equation}
		\frac{\Gamma ({\bm r})}{\Gamma^{(0)}} = \frac{6\pi c}{\omega_0} \textrm{Im} \{ {\bm n} \cdot {\mathds G} ({\bm r}, {\bm r}; \omega_0 ) \cdot {\bm n} \}\, ,
\label{TaxaEmissaoEspontanea2}
\end{equation}
where ${\bm n} = {\bm d}/|{\bm d} |$, and $\mathds{G} ({\bm r}, {\bm r}'; \omega)$ is the dyadic Green function of the system, which encodes the influence of surrounding bodies on the emitter's lifetime. It satisfies 
\begin{equation}
	\nabla \times \nabla \times \mathds{G} ({\bm r}, {\bm r^{\prime}}; \omega ) - \frac{\omega^2}{c^2} \, \mathds{G} ({\bm r}, {\bm r^{\prime}}; \omega) = \mathds{I} \delta ({\bm r} - {\bm r^{\prime}}).
\label{FuncaoGreen}
\end{equation}
\noindent along with proper BC.

Plugging the information from Green’s function associated with the electromagnetic field generated by the oscillating dipole source and scattered (reflected) by a planar surface, it can be shown that the perpendicular and parallel contributions are~\cite{Szilard-2016, Kort-Kamp-2015}
\begin{align}
    \frac{\Gamma_{\perp}}{\Gamma^{(0)}} &= \frac{d_{\perp}^2}{|{\bm d}|^2} \left\{ 1 + \frac{3}{2} \int_0^{\infty} dk_\parallel \frac{k_{\parallel}^3}{k_{0}^3 \xi} \textrm{Re}\left[ r_{pp} \, e^{2i\xi z} \right] \right\}\, , \label{GammaPerp} \\
    \frac{\Gamma_{\parallel}}{\Gamma^{(0)}} &= \frac{d_{\parallel}^2}{|{\bm d}|^2} \left\{ 1 + \frac{3}{4} \int_0^{\infty} dk_\parallel \frac{k_{\parallel}}{k_{0}^3 \xi} \textrm{Re}\left[\left(k_0^2 r_{ss} - \xi^2 r_{pp} \right) e^{2i\xi z}\right] \right\}\, , \label{GammaPar}
\end{align}
\noindent where $\xi = \sqrt{k_0^2 - k_{\parallel}^2}$, and $r_{ss}$ and $r_{pp}$ are the usual Fresnel reflection coefficients for a flat interface between vacuum and a homogeneous medium \cite{Novotny-Book}
\begin{align}
    r_{ss} = \frac{{k_z}_0 - {k_z}_1}{{k_z}_0 + {k_z}_1} \;\;\;\; {\rm and} \;\;\;\;
    r_{pp} = \frac{\varepsilon_e {k_z}_0 - {k_z}_1}{\varepsilon_e{k_z}_0 +{k_z}_1}\, , \label{FresnelCoefficients}
\end{align}
\noindent where $\varepsilon_e$ is the effective dielectric constant of the medium, and $k_{z1} = \sqrt{\varepsilon_e k_0^2 - k_{\parallel}^2}$.

\begin{figure}[t]
  \centering
 \includegraphics[width=0.74\linewidth]{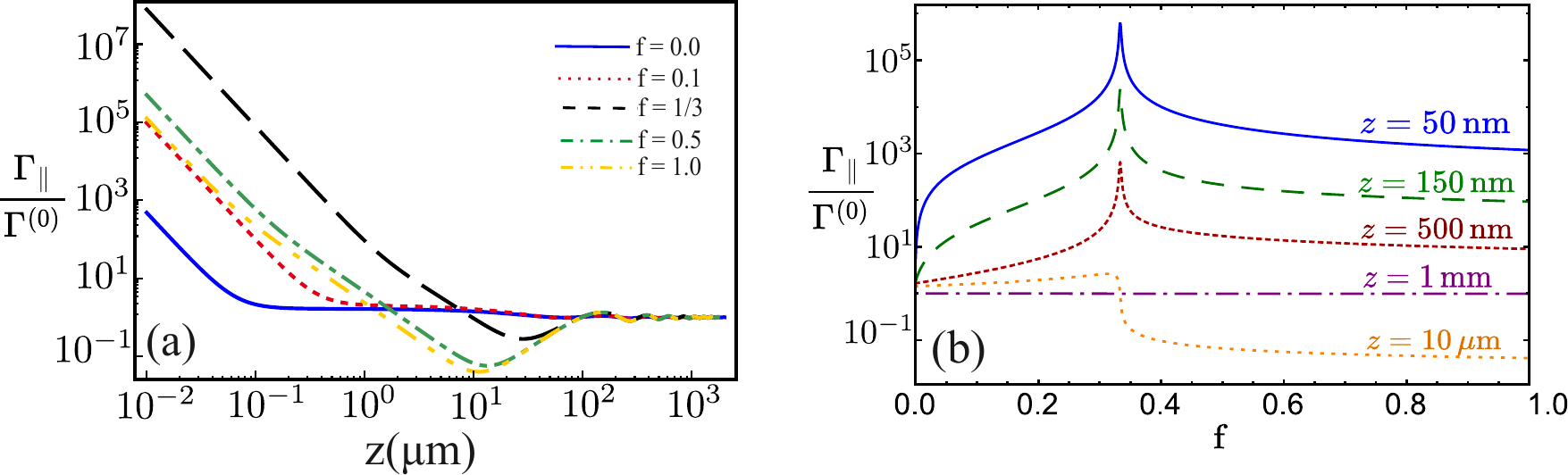}
  \caption{Normalized SE rate $\Gamma_\parallel/\Gamma^{(0)}$ as a function of (a) $z$ for different filling fractions $f$ and (b) $f$ for different distances $z$. In both cases, the composite medium consists of gold spherical inclusions. Adapted from Ref.~\cite{Szilard-2016}.}
  \label{PercolationResults}
\end{figure}

Figure~\ref{PercolationResults} computes the SE rate as a function of the distance $z$ for various filling fractions $f$ [Fig.~\ref{PercolationResults}(a)] and as a function of $f$ for various values of $z$ [Fig.~\ref{PercolationResults}(b)]. According to BEMT for spherical inclusions, the composite medium is metal-like for $f > 1/3$ and dielectric-like for $f < 1/3$~\cite{Sahimi-1993}. The emitter, a cesium atom with $\lambda_0 = 450$ $\mu$m, experiences a significant SE rate enhancement in the presence of the composite medium, mainly compared to homogeneous cases ($f = 0, 1$). For both parallel and perpendicular dipole orientations, the SE rate can increase by five to six orders of magnitude at $z \sim 100$ nm.

Figure~\ref{PercolationResults} also illustrates how the transition from far-field to near-field effects depends on the filling fraction. Interestingly, at the BEMT percolation threshold for spherical inclusions ${f = f_c = 1/3}$, near-field effects become noticeable at distances $z \lesssim  1$ $\mu$m. At such distances, light emission is strongly influenced by evanescent electromagnetic modes ($k_{\parallel}>k_0$), which are confined near the air-substrate interface. In the extreme near-field regime, an approximate analytic expression for the SE rate can be obtained by taking the quasi-static limit ($c\to\infty$) in Eqs.~(\ref{GammaPerp}) and (\ref{GammaPar}), leading to
\begin{equation}
\frac{\Gamma_{\perp}}{\Gamma^{(0)}} \simeq 2 \frac{\Gamma_{\parallel}}{\Gamma^{(0)}} \simeq \dfrac{3}{4} \dfrac{1}{z^3} \dfrac{\textrm{Im}[\varepsilon_e]}{|\varepsilon_e+1|^2}\, .
\label{SENF}
\end{equation}

\noindent This equation reveals that the SE rate initially increases with Im$[\varepsilon_{e}]$ for small $f$, due to enhanced absorption. However, it decreases at higher $f$ as the medium becomes more metallic, scaling as 1/Im$[\varepsilon_{e}] \ll 1$. The SE rate reaches its maximum precisely at ${f_{c} = 1/3}$, which corresponds to the percolation transition predicted by BEMT for spherical inclusions (see Fig.~\ref{PercolationResults}).  A similar behavior holds for the perpendicular dipole contribution. We also verified that these findings are robust against variations in the shape of the inclusions, the choice of effective medium theory, and across a broad spectral range of the emission frequency~\cite{Szilard-2016}.

The increased SE rate in composite media is tied to universal properties of the percolation phase transition, particularly the enhanced, scale-invariant fluctuations of current and electric fields near the percolation critical point. These fluctuations modify the electromagnetic modes, directly affecting the SE. Close to the percolation threshold, composite media exhibit highly localized, subwavelength-confined resonant plasmon excitations, which generate large spatial fluctuations in the electromagnetic field intensity~\cite{Sarychev2000}.

\subsection{Atom near a VO$_2$ slab: hysteresis in SE rate}

\begin{figure}[t]
  \centering
 \includegraphics[width=0.98\linewidth]{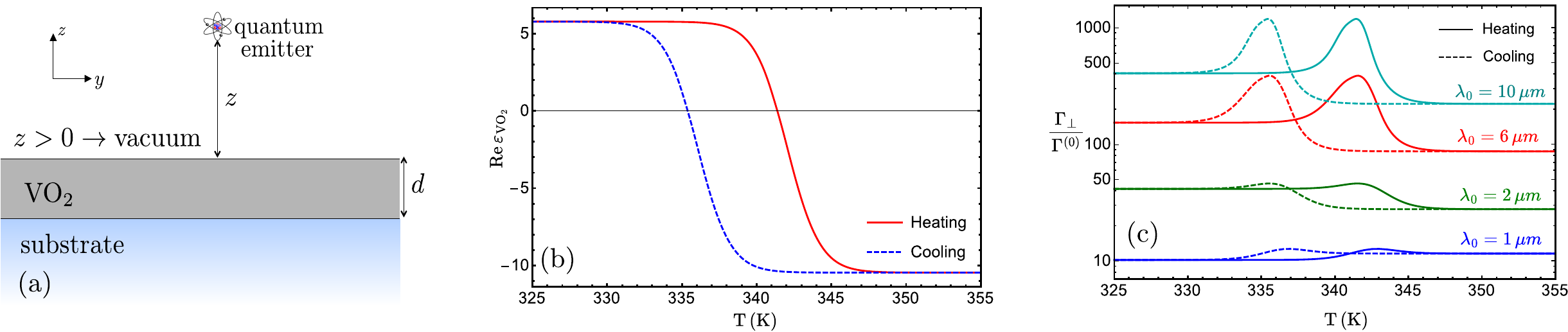}
  \caption{(a) System under study. (b) Dielectric constant of VO$_2$ at $\lambda_0 = 3$ $\mu$m. (c) SE rate $\Gamma_\perp/\Gamma^{(0)}$ as a function of $T$ for different $\lambda_0$. We set $d = 200$ nm. Adapted from Ref.~\cite{Szilard-2019}.}
  \label{SE-VO2}
\end{figure}

Effective medium techniques also enable modeling phase transitions in materials beyond composite media. Here, we use BEMT to investigate the SE of a quantum emitter near a vanadium dioxide (VO$_2$) thin film of thickness $d$ grown on a sapphire substrate, as in Fig.~\ref{SE-VO2}(a) \cite{Szilard-2019, tese-Dani}. VO$_2$ is notable for undergoing a first-order metal-insulator transition (MIT) at a relatively low critical temperature $T_c \sim 68^\circ$C \cite{Morin-1959}, along with a significant increase in its electrical conductivity up to five orders of magnitude \cite{Verleur-1968}.

Recently, VO$_2$ has gained attention for its potential technological applications~\cite{Yang-2015,  Pergament-2013, KortKamp-2018, Jin-2024} and its thermal hysteresis in optical properties~\cite{Peterseim-2016}. Figure~\ref{SE-VO2}(b) illustrates its effective dielectric constant at $\lambda_0 = 3$ $\mu$m, highlighting the hysteresis during the MIT. The thermal hysteresis manifests in various quantum vacuum fluctuation effects~\cite{Cueff-2015, Cueff-2020, Cunningham-2021, Liu-2023}, including the Purcell effect~\cite{Szilard-2019, tese-Dani}.

Figure~\ref{SE-VO2}(c) shows the SE rate in the perpendicular configuration as a function of temperature for different values of $\lambda_0$, with an electric emitter positioned at $z = 50$ nm from the VO$_2$ film. The SE rate varies non-monotonically with temperature as VO$_2$ transitions across the MIT region. The thermal hysteresis appears in the SE rate: the temperature at which the maximum decay rate occurs depends on the system's heating or cooling history. Additionally, the SE rate peak is more pronounced for longer wavelengths, where material absorption dominates over other decay mechanisms, like plasmon excitations and single-photon emission~\cite{Szilard-2016}. This SE rate peak increases by up to a factor of $10^3$ compared to the free-space value for an emitter with $\lambda_0 \lesssim 10$ $\mu$m. Similar results are observed for the parallel contribution. In Ref.~\cite{Szilard-2019}, the SE rate of magnetic emitters was also investigated, with the magnetic SE peaks occurring in the shorter wavelength range.

\subsection{Controlling SE with phosphorene under mechanical strain}

\begin{figure}[b]
    \centering    \includegraphics[width=0.34\linewidth]{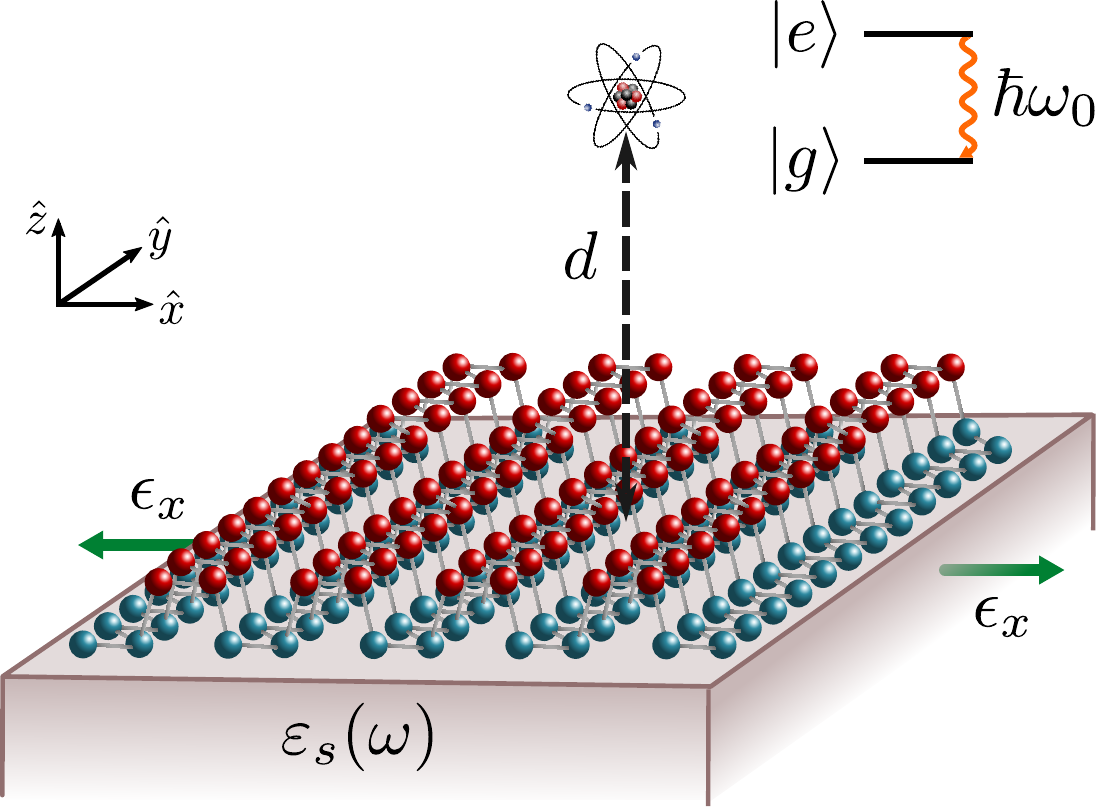}
    \caption{Quantum emitter close to a phosphorene/SiC medium, subjected to uniaxial strain (applied along the $x$-direction, in this figure). From Ref.~\cite{Abrantes2023}.}
    \label{FigSEPhosphorene1}
\end{figure}

As our last application, we discuss the SE of an emitter near a phosphorene sheet grown on a silicon carbide (SiC) substrate, as illustrated in Fig.~\ref{FigSEPhosphorene1} \cite{Abrantes2023}. The electronic structure of phosphorene is highly sensitive to strain due to its puckered lattice and flexibility \cite{Peeters-StrainPhosphorene-Model, Midtvedt-Lewenkopf-Croy-JPCM, Midtvedt-Lewenkopf-Croy-2DMat}, which allows it to sustain up to $30\%$ strain \cite{Flexibility-Phosphorene-1, Flexibility-Phosphorene-2} within state-of-the-art experiments \cite{Exp-Strain-Ph-1, Exp-Strain-Ph-2}. Consequently, the electronic properties of phosphorene and its anisotropic character are altered, affecting its optical response and plasmonic characteristics \cite{Nemilentsau-Hanson-PRL-2016}. This approach provides a clear advantage when compared to previous proposals based on electromagnetic fields acting as external agents. For instance, in Ref.~\cite{Kort-Kamp-2015}, the authors investigate the SE of an emitter near a graphene-coated substrate subjected to an external magnetic field, revealing significant variations in the SE rate even at moderate field intensities. However, the external magnetic field may notably alter the atomic resonance frequency through the Zeeman effect, a factor that must be considered in experimental setups. In contrast, mechanical strain does not introduce such complications.

We model the quantum emitter as a two-level system with an energy difference $\hbar\omega_0 = h c/\lambda_0$, where the transition occurs between excited $\ket{e}$ and ground $\ket{g}$ states. The emitter is located at a distance $d$ from a phosphorene sheet. This section focuses on electric dipole transitions, though the Purcell effect for magnetic dipole transitions was also explored in Ref.~\cite{Abrantes2023}. Results for uniaxial strain along the $y$-axis (zigzag) are presented, but similar analyses were also conducted for strain along the $x$-axis (armchair).

Following previous discussions, the SE rate is calculated from Eq.~(\ref{TaxaEmissaoEspontanea2}), by writing $\mathds{G} (\bm{r},\bm{r}, \omega)$ in terms of the diagonal components of the reflection matrices ($r_{ss}$ and $r_{pp}$) \cite{Kort-Kamp-2015}. These, in turn, can be calculated by solving the Maxwell equations with suitable BCs, accounting for the optical conductivity of phosphorene and the electrical permittivity of the substrate \cite{Abrantes2023}. The expressions corresponding to transition dipole moments parallel to the $x$ (armchair), $y$ (zigzag), and $z$ (perpendicular) directions, respectively, become
\begin{align}
	\frac{\Gamma_{x}}{\Gamma^{(0)}} &= 1 + \frac{3}{4\pi k_0} \, {\rm Im} \left[ i \int d^2 \bm{k}_{\parallel} \frac{e^{2 i \sqrt{k_0^2 - k^2_{\parallel}} d}}{ k^2_{\parallel} \sqrt{k_0^2 - k^2_{\parallel}}} \left( k^2_y r_{ss} - \frac{k^2_x (k_0^2 - k^2_{\parallel})}{k^2_0} r_{pp} \right) \right], \\
	\frac{\Gamma_{y}}{\Gamma^{(0)}} &= 1 + \frac{3}{4\pi k_0} \, {\rm Im} \left[ i \int d^2 \bm{k}_{\parallel} \frac{e^{2 i \sqrt{k_0^2 - k^2_{\parallel}} d}}{k^2_{\parallel} \sqrt{k_0^2 - k^2_{\parallel}}} \left( k^2_x r_{ss} - \frac{k^2_y (k_0^2 - k^2_{\parallel})}{k^2_0} r_{pp} \right) \right], \\
	\frac{\Gamma_{z}}{\Gamma^{(0)}} &= 1 +  \frac{3}{4\pi k_0^3} \, {\rm Im} \left[ i \int d^2 \bm{k}_{\parallel} \frac{k^2_{\parallel} \, e^{2 i \sqrt{k_0^2 - k^2_{\parallel}} d}}{\sqrt{k_0^2 - k^2_{\parallel}}} r_{pp} \right].  \label{Gammazz}
\end{align}
\noindent Due to the anisotropic nature of phosphorene, $\Gamma_{x} \neq \Gamma_{y}$, unlike the examples discussed in the previous sections. For our analysis, we set the Fermi energy of phosphorene at $E_{\rm F}=0.7$ eV \cite{Das-Roelofs-ACS-Nano-2014} and consider emitters with $\lambda_0 = 4.1, 10$ $\mu$m \cite{Review-QuantumDots, Hulet-85}.

\begin{figure}[t]
    \centering    \includegraphics[width=1\linewidth]{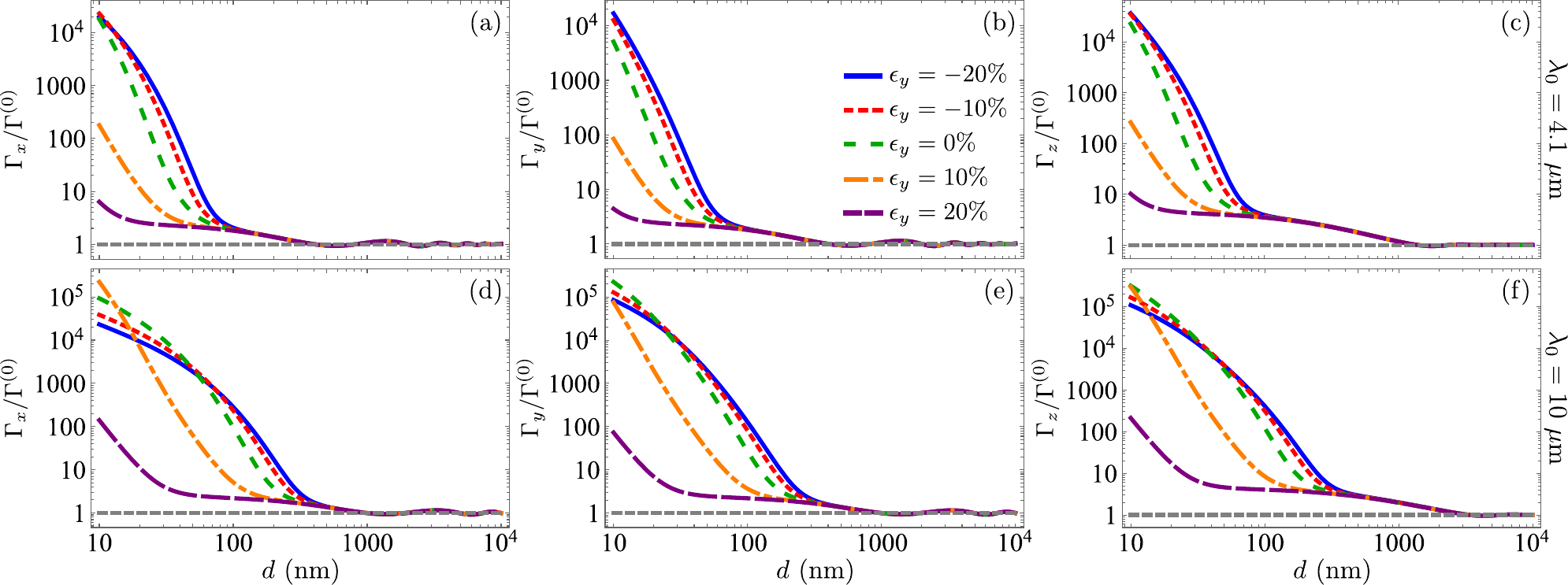}
    \caption{SE rates as functions of the distance $d$, considering uniaxial strain $\epsilon_y$ with different intensities, and transition wavelengths (a)-(c) $4.1$ $\mu$m, and (d)-(f) $10$ $\mu$m. Adapted from Ref.~\cite{Abrantes2023}.}
    \label{FigSEPhosphorene2}
\end{figure}

Figure \ref{FigSEPhosphorene2} shows how the SE rates vary with distance $d$ for different strain values $\epsilon_y$. Note that all SE rates approach the free-space value at large distances ($d \gg \lambda_0$), regardless of strain, since the influence of phosphorene becomes progressively less relevant. In the short-distance regime, significant changes in SE rates are observed even for relaxed sheets ($\epsilon_y = 0\%$). However, when strain is applied, the SE rates may be substantially modulated. When $\lambda_0 = 4.1$ $\mu$m, compressive (tensile) strain $\epsilon_y<0$ ($\epsilon_y>0$) enhances (decreases) the SE rates, which can be traced back to the behavior of the intraband and interband contributions of the optical conductivities of strained phosphorene \cite{Abrantes2023}. Conversely, when $\lambda_0 = 10$ $\mu$m, a non-monotonic behavior of the SE rates with respect to strain takes place. 

\begin{figure}[b!]
    \centering    \includegraphics[width=0.58\linewidth]{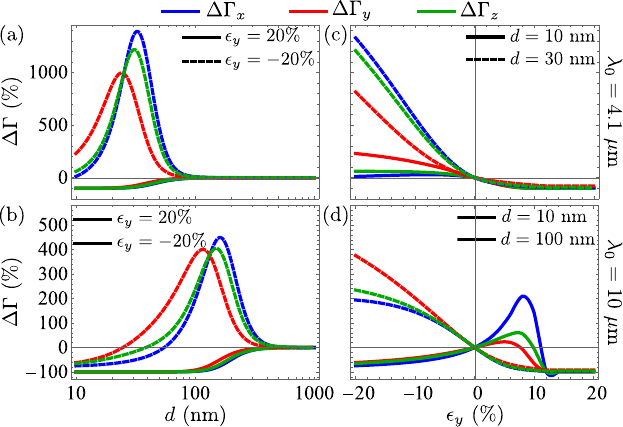}
    \caption{Percentage variation in the electric SE rates as functions of (a)-(b) $d$ and (c)-(d) $\epsilon_y$. Adapted from Ref.~\cite{Abrantes2023}.}
    \label{FigSEPhosphorene3}
\end{figure}

To quantify the impact of strain on SE, we define the quantity $\Delta \Gamma_{\nu} = [\Gamma_{\nu} (\epsilon_{y}\neq 0) - \Gamma_{\nu} (\epsilon_{y}=0)]/\Gamma_{\nu} ({\epsilon_{y}=0})$, where $\nu = \{x, y, z\}$ and $\Gamma_{\nu}(\epsilon_{y}\neq 0)$ [$\Gamma_{\nu} ({\epsilon_{y} = 0})$] stands for the SE rate near strained (relaxed) phosphorene. In Fig.~\ref{FigSEPhosphorene3}, we exhibit the percentage variation in SE rates for both values of $\lambda_0$. Figures~\ref{FigSEPhosphorene3}(a) and (b) show these variations as functions of $d$, while Figs.~\ref{FigSEPhosphorene3}(c) and (d) are calculations as functions of $\epsilon_y$. In all results, $\Delta \Gamma_{x} \neq \Delta \Gamma_{y}$ highlights phosphorene's anisotropic nature. For $\lambda_0 = 4.1$ $\mu$m, compressive strain can remarkably enhance the  Purcell effect up to $1300 \%$, whereas tensile strain nearly suppresses the Purcell effect for both values of $\lambda_0$. Ultimately, these plots underscore that strain can effectively switch on and off quantum emission on demand. 

Lastly, let us discuss the decay channels of the emitted quanta. Without any surrounding media, a quantum emitter undergoes a radiative decay, emitting propagating (Prop) modes detectable in the far-field. However, when near some environment, other decay channels may arise and become increasingly relevant in the near-field \cite{Kort-Kamp-2015, Szilard-2016}. The emitted photon can generate total internal reflection (TIR) modes that propagate within the substrate and evanesce in vacuum if losses are insignificant. They emerge for $k_0 < k_{\parallel} < n_s k_0$, with the medium refraction index $n_s = {\rm Re} \left[ \sqrt{\varepsilon_s/\varepsilon_0} \right]$. In addition, the emitter may decay nonradiatively when $k_{\parallel} > n_s k_0$, originating lossy surface waves (LSWs), in which case the excitation energy is transferred to the medium, but it is eventually damped and dissipated as heat. From Eq. (\ref{Gammazz}), we can obtain the contributions of each channel to the SE rate as \cite{Kort-Kamp-2015, Szilard-2016}
\begin{align}
	\frac{\Gamma_{z,{\rm Prop}}}{\Gamma^{(0)}} &\simeq 1 + \frac{3}{4 \pi k^3_0} \int_0^{k_0} \!\! d k_{\parallel} \int_0^{2\pi} \!\! d\phi \frac{k^3_{\parallel} \, {\rm Re} \left[ e^{2 i \sqrt{k_0^2 - k^2_{\parallel}}d} r_{pp} \right]}{ \sqrt{k_0^2 - k^2_{\parallel}}} ,\\
	\frac{\Gamma_{z,{\rm TIR}}}{\Gamma^{(0)}} &\simeq \frac{3}{4\pi k^3_0} \int_{k_0}^{n_s k_0} \!\! dk_{\parallel} \int_0^{2\pi} \!\! d\phi \frac{k^3_{\parallel} e^{-2 \sqrt{k^2_{\parallel} - k_0^2} d} \, {\rm Im} \left[r_{pp}\right]}{\sqrt{k^2_{\parallel} - k_0^2}}, \\
	\frac{\Gamma_{z,{\rm LSW}}}{\Gamma^{(0)}} &\simeq \frac{3}{4\pi k^3_0} \int_{n_s k_0}^{\infty} \!\! d k_{\parallel} \int_0^{2\pi} \!\! d\phi \frac{k^3_{\parallel} e^{- 2 \sqrt{k^2_{\parallel} - k_0^2} d} \, {\rm Im} \left[r_{pp}\right]}{\sqrt{k^2_{\parallel} - k_0^2}}. 
\end{align}
\noindent The probabilities $p_{z,{\rm Prop}}$, $p_{z,{\rm TIR}}$, and $p_{z,{\rm LSW}}$ of energy emission in each decay channel are calculated from the ratio between the partial and total SE rates.

\begin{figure}[b!]
    \centering    \includegraphics[width=0.58\linewidth]{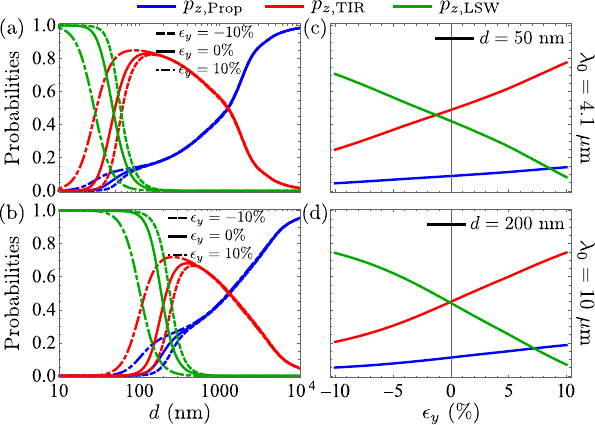}
    \caption{Decay channels probabilities as functions of (a)-(b) $d$ and (c)-(d) $\epsilon_y$. Adapted from Ref.~\cite{Abrantes2023}.}
    \label{FigSEPhosphorene4}
\end{figure}

Figure \ref{FigSEPhosphorene4} shows the decay channel probabilities for both values of $\lambda_0$. In Figs.~\ref{FigSEPhosphorene4}(a) and \ref{FigSEPhosphorene4}(b), the decay channels are analyzed as functions of the distance $d$ for different strain values. As the emitter moves further from the phosphorene sheet, the influence of this material is reduced, and Prop modes become dominant. However, competition between the TIR and LSW modes arises as $d$ decreases. At this point, we can see how versatile uniform strain can be in enabling control over which mode dominates the emission pathway. For even smaller values of $d$ (near-field regime), LSW modes fully dominate the SE process.  Figures~\ref{FigSEPhosphorene4}(c) and \ref{FigSEPhosphorene4}(d) illustrate the decay probabilities as functions of strain for a fixed distance $d$ in the intermediate regime, highlighting that the strain provides a convenient platform to harness the different decay channels.

%
%
%
%
\section{Final remarks and conclusions}

In this work, we explored several quantum vacuum effects within the framework of QED and provided a comprehensive historical overview of phenomena such as SE and the Purcell effect. We then proceeded to examine specific examples where the environment surrounding an atomic system plays a crucial role in significantly modifying the SE rate. In terms of practical applications, some of these approaches are more experimentally feasible than others. For instance, applying a magnetic field to a graphene sheet or introducing mechanical strain to phosphorene appears to be more straightforward experimentally than manipulating the concentration of metallic inclusions within a dielectric matrix.

However, SE is far from the only quantum phenomenon that can be influenced by the surrounding media. Numerous other phenomena, also described within the framework of QED, such as TPSE \cite{Yuri-Tese, Hayat2008, Nevet2010, Poddubny2012, Rivera2017, Yuri-2019, Yuri-2020, Yuri-2022, Weitzel-2022, Whisler2023}, dispersive interactions \cite{Cysne-CasimirPolder-PRA-2014, Jiang2019, Fuchs2018, Haug2019, Silvestre-QR-PRA-2019, Abrantes-QR-PRB-2021, Passante-2020, tese-Patricia, Abrantes-DI-PRA-2021, BSLu-Universe-2021, Laliotis-AVSQSci-2021, Rodriguez-Lopez-NatCommun-2017, Muniz-Farina-Kort-Kamp-2021, Kilianski2024}, near-field radiative heat transfer \cite{Ekeroth, Ghanekar, NFRHTGraphene-PRApp, NFRHTGraphene-PRB, Pascale2023, Correa2024, Laura2022, Mittapally2023, Salihoglu2023}, and resonance energy transfer \cite{tese-Patricia, Abrantes-RET-2020, Abrantes-RET-2021, Lee2023, Liu2023, Nayem2023, Joao-RET-2025}, can also be controlled, and their manipulation holds substantial potential for real-world applications. In this context, developing techniques to govern radiation-matter interactions at the nanoscale emerges as an exceptionally promising research direction, with direct implications for fields such as nano-optics and plasmonics. By advancing our ability to control these interactions precisely, we open the door to novel applications and innovations in these rapidly evolving areas of technology.

\section*{Acknowledgment}

The authors acknowledge W. J. M. Kort-Kamp for discussions. C.F. acknowledges funding from FAPDF (Funda\c{c}\~{a}o de Apoio \`{a} Pesquisa do Distrito Federal, Grants No. 308641/2022-1, 204.376/2024, and 00193-00001817/2023-43), CNPq (Conselho Nacional de Desenvolvimento Cient\'\i fico e Tecnol\'ogico, Grants No. 308641/2022-1 and 408735/2023-6), and FAPERJ (Funda\c c\~ao Carlos Chagas Filho de Amparo \`a Pesquisa do Estado do Rio de Janeiro, Grant 204.376/2024). P.P.A. acknowledges funding from CNPq (Grant No. 152050/2024-8).

\renewcommand{\bibname}{References}
\begingroup
\let\cleardoublepage\relax

\endgroup

\end{document}